\documentclass[aps,pra,superscriptaddress,epsfigure,twocolumn]{revtex4-1}
\usepackage{graphicx}
\usepackage{amsmath}    
\usepackage{latexsym}
\usepackage{amsfonts}   
\usepackage{amssymb}
\usepackage{array}      
\usepackage{epsfig}
\usepackage{txfonts}
\usepackage{hyperref}
\usepackage{color}
\usepackage{braket}
\usepackage{float}
\usepackage{mathrsfs}
\usepackage{mathtools}

\begin{document}
%
%--------------------------------------------------------------------------------------------------------------------------------
\title{Three-qubit refrigerator with two-body interactions}
\author{Adam Hewgill}
\affiliation{Centre  for  Theoretical  Atomic,  Molecular  and  Optical  Physics, Queen's  University  Belfast,  Belfast  BT7 1NN,  United  Kingdom}
\author{J. Onam Gonz\'{a}lez}
\author{Jos\'{e} P. Palao}
\author{Daniel Alonso}
\affiliation{Dpto. de F\'{i}sica and IUdEA: Instituto Universitario de Estudios Avanzados, Universidad de La Laguna, 38203 Spain}

\author{Alessandro Ferraro}
\affiliation{Centre  for  Theoretical  Atomic,  Molecular  and  Optical  Physics, Queen's  University  Belfast,  Belfast  BT7 1NN,  United  Kingdom}
\author{Gabriele De Chiara}
\affiliation{Centre  for  Theoretical  Atomic,  Molecular  and  Optical  Physics, Queen's  University  Belfast,  Belfast  BT7 1NN,  United  Kingdom}
\affiliation{Laboratoire Charles Coulomb (L2C), UMR 5221 CNRS-Universit\'e de Montpellier, F-34095 Montpellier, France}

\begin{abstract}
We propose a three-qubit setup for the implementation of a variety of quantum thermal machines where all heat fluxes and work production can be controlled. An important configuration that can be designed is that of an absorption refrigerator, extracting heat from the coldest reservoir without the need of external work supply. Remarkably, we achieve this regime by using only two-body interactions instead of the widely employed three-body interactions. This configuration could be more easily realised in current experimental setups. We model the open-system dynamics with both a global and a local master equation thermodynamic-consistent approach. Finally, we show how this model can be employed as a heat valve, in which by varying the local field of one of the two qubits allows one to control and amplify the heat current between the other qubits.
\end{abstract}

\maketitle
%--------------------------------------------------------------------------------------------------------------------------------
%
%--------------------------------------------------------------------------------------------------------------------------------
\section{Introduction}

Quantum thermodynamics studies the emergence of thermodynamic behavior in systems that are quantum in nature \cite{Binder2018,Kosloff2013,Vinjanampathy2016}. One of the main aims of thermodynamics is the design and realisation of thermal machines, devices relying on thermodynamic processes to achieve some desired objective. Besides, the advances in controlling quantum systems has allowed to characterize thermal machines in the quantum domain~\cite{Vinjanampathy2016,GooldReview,XuerebReview,Binder2018,Mitchison2019}. Quantum absorption machines have recently sparked a wide interest since they can perform the desired task without the input of any external source of work. These machines use heat to power themselves and therefore require at least three different thermal reservoirs. The main focus of research into absorption machines has been in coming up with quantum absorption refrigerators \cite{Palao2001,LindenPRL2010,Levy2012b,LevyPRE2012,CorreaPRE2013,VenturelliPRL2013,CorreaPRE2014,CorreaSR2014,YuPRE2014,LeggioPRA2015,SilvaPRE2015,MitchisonQST2016,SilvaPRE2016,HoferPRB2016,DoyeuxPRE2016,MuNJP2017,NimmrichterQ2017,RouletPRE2017,ManPRE2017,FogedbyEPL2018,ErdmanPRB2018,Holubec2018,DuNJP2018,HartlePRB2018,SegalPRE2018,KilgourPRE2018,SeahPRE2018,Maslennikov19,DasEPL2019,manzano2019boosting}, which remove energy from a cold reservoir and dump it into a hot reservoir using a third, hotter work reservoir. 

In the seminal work of Linden and coworkers \cite{LindenPRL2010}, an absorption refrigerator was proposed which employs three qubits as a working substance, each connected to a thermal reservoir at different temperatures. In this model, the three qubits interact via a three-body coupling. An experimental realisation of a similar model with trapped ions exploiting three-body interactions has been reported in \cite{Maslennikov19}.

For continuous-variable systems, e.g. quantum oscillators, it has been rigorously proven by Martinez and Paz that an autonomous absorption refrigerator cannot exist with only time-independent quadratic Hamiltonians and that some form of nonlinearity would be required~\cite{MartinezPRL2013}. For discrete quantum systems, the minimum requirements on the form of the Hamiltonian are not known and the necessity of three-body interactions remains an open problem in general. 

The authors of Ref.~ \cite{LindenPRL2010}, well aware of this issue, described a modified model involving one qubit and one qutrit with a suitably engineered interaction that do not require three-body interactions. To the best of our knowledge, however, the question whether an absorption refrigerator, i.e. not requiring work, could be designed with three qubits interacting with two-body spin-spin couplings has not been settled. This is the objective of this work: we demonstrate that simple two-spin exchange interactions among three qubits connected to three thermal reservoirs are enough to refrigerate the coldest reservoir without the expense of any external work.

Any investigation of the thermodynamic properties will inevitably lead to considering systems that have some connections to external environments. The exchange of energy between the system and the environments can be described using the theory of open quantum systems \cite{Breuer2002}. There are several methods to model the system-environment couplings, which are in turn dependent on the physical situation under consideration. In particular, the Gorini-Kossakowski-Lindblad-Sudarshan (GKLS) master equation describes the time evolution of a system that is weakly connected to a Markovian environment \cite{GSK76,lindblad1976}. The main ingredients in the GKLS equation are the jump operators, which describe the transitions assisted by the thermal environments. The specific jump operators in each case depend on the physical model for the couplings and the environments.

When the couplings between the system and the environments result from repeated short interactions, as for example in collisional models, the jump operators are related only to the subsystem directly in contact with each thermal bath \cite{Scarani2002,Ziman2002,Karevski2009,Giovannetti2012,Landi2014,Barra2015,StrasbergPRX2016,ManzanoPRE2016,Shammah2018,SeahPRE2018,Ciccarello2017,PezzuttoNJP2016,pezzutto2018out,cusumano2018entropy,CampbellPRA2018,GrossQST2018,rodrigues2019thermodynamics,seah2019collisional,SeahPRE2019,strasberg2019repeated,ManPRA2019,ManQIP2019,ehrich2019micro}. In this case the derivation of the GKLS master equation is frequently termed as the {\it local} (or boundary-driven) approach. The external work required to switch on and off the system bath interaction guarantees the thermodynamic consistency of the model \cite{Barra2015,StrasbergPRX2016,De_Chiara_2018,Kosloff2019}. A different situation occurs when considering harmonic baths permanently coupled to the system. Assuming in addition that the system inner time scales are smaller than the characteristic time of the relaxation induced by the thermal baths, the secular (or rotating wave) approximation applies \cite{Breuer2002}. Thus, the dynamics is again described by a master equation in GKLS form. However, now the jump operators are related in general to the entire system. The corresponding derivation of the master equation is sometimes termed as the {\it global} approach. The use of the terms {\it local} and {\it global} could be misleading, since the {\it local} approach may lead to a good approximation to the exact dynamics for systems coupled to harmonic baths in certain parameter regimes for which the secular approximation fails \cite{HoferNJP2017,Gonzalez2017}. However, for time-independent models thermodynamic consistency may not be ensured \cite{LevyKosloffEPL2014,Kosloff2019}. Remarkably, for time-dependent setups a detailed thermodynamic analysis may overcome this limitation \cite{Barra2015,StrasbergPRX2016,De_Chiara_2018}. 

Within this work we will study a system of three qubits whose couplings with the environment can be described either by the repeated interaction model (local master equation) or by the harmonic model (global master equation). We will show that the differences between the corresponding master equations will lead to distinct behavior in the thermodynamic functioning of the system. 
%In particular, we will analyse the smallest possible quantum implementation of an absorption refrigerator, consisting in three qubits \cite{LindenPRL2010}. Previous research considers mainly three-qubit models with a three-body internal coupling continuously connected to bosonic reservoirs. This choice may be motivated by the analogy with harmonic models for absorption refrigerators. For this sort of models, it have been shown that it is impossible to build a quantum absorption refrigerator based on linear internal couplings \cite{Martinez2013}. However, a three body coupling could be extremely involved experimentally~\cite{Maslennikov19}. 
We will perform an exhaustive exploration of the parameter space to characterize all the possible operating regimes of the system. We demonstrate that in both models an absorption refrigerator mode can be found without the need of three-body interactions.

The paper is organized as follows: in Sec. \ref{sec:3Qmodel} we introduce the three-qubit model with two-body internal couplings. We also briefly describe the derivation of the corresponding master equation when considering either a continuous coupling to harmonic baths or a repeated interaction model. Particular attention is paid to the definition of the heat currents and the thermodynamic consistency of each description. In Sec \ref{sec:modes} we explore the possible operating modes of the system in each case, with special attention to absorption refrigerators. In Sec. \ref{sec:valve} we analyse the limitations of the three-qubit model when working as a heat valve. Finally, we summarize our findings in Sec. \ref{sec:conclusions}. In Appendix~\ref{sec:corre} we explore the quantum correlations between the different components of the device. 

%--------------------------------------------------------------------------------------------------------------------------------
%
%--------------------------------------------------------------------------------------------------------------------------------

\section{Three-qubit device}\label{sec:3Qmodel}

We consider three qubits that are coupled to each other via an XXZ Hamiltonian:
\begin{eqnarray} \label{eq:HS}
\hat{H}_S&=& \hat{H}_I+\sum_{i=1}^3 \hat{H}_{L_{i}},
\\
\hat{H}_{L_{i}}&=& \hbar B_i \, \hat{\sigma}_z^{i},\label{eq:HL}
\\
\hat{H}_I &=& \sum_{\substack{i,j \\ j>i} } \hbar J_{ij} \,  (\, \hat{\sigma}_x^{i}\, \hat{\sigma}_x^{j}+\hat{\sigma}_y^{i}\hat{\sigma}_y^{j}\, )+ \hbar \Delta_{ij} \, \hat{\sigma}_z^{i}\, \hat{\sigma}_z^{j},
\end{eqnarray} 
where $\hat{\sigma}_{x,y,z}^{i}$ are the Pauli spin operators, $B_i, J_{ij}$ and $\Delta_{ij}$ are the values of the local magnetic field, the qubit coupling strength and the interaction anisotropy, respectively. Without loss of generality we assume them to be positive coefficients. Besides, it can be easily shown that the total magnetization $\hat{\sigma}_z\equiv \hat{\sigma}_z^{1}+\hat{\sigma}_z^{2}+\hat{\sigma}_z^3$ is a conserved quantity with respect to the system Hamiltonian
\begin{equation} \label{eq:conserved}
[\hat{H}_S,\hat{\sigma}_z]=0 \,.
\end{equation}

In the following subsections we describe two microscopic models for the interaction of the system with the environment. Although the XXZ model is quite common in the literature, it is not the most general form of two-qubit interaction. Nevertheless, as we show in the following, this Hamiltonian leads to a significant level of control in the functioning of the system as a thermal device.

\subsection{Harmonic-bath model} \label{global}
In the harmonic-bath model, the quantum system is permanently connected to thermal baths. We consider three bosonic reservoirs $R_1$, $R_2$ and $R_3$ at temperatures $T_1$, $T_2$ and $T_3>T_2>T_1$. Each qubit $i$ is attached to the bath $R_i$ with a well-defined temperature $T_i$. The free environments Hamiltonians read:
\begin{equation} \label{eq:HR}
\hat{H}_{R_i}=\sum_{\mu} \hbar \, \omega_{\mu,i} \, \hat{b}_{\mu,i}^\dagger \, \hat{b}_{\mu,i} \,, \ \ (i=1,2,3) \,,
\end{equation}
being $\hat{b}_{\mu,i}^\dagger$ and $\hat{b}_{\mu,i}$ the creation and annihilation operators corresponding to the mode with frequency $\omega_{\mu,i}$. Besides, the interaction between the system and each reservoir is described by:
\begin{equation} \label{eq:HSR}
\hat{H}_{SR_i}= \hat{\sigma}_x^{i} \otimes \hat{R}^{i} \,,
\end{equation}
with
\begin{equation} \label{eq:B}
\hat{R}^{i} =\hbar \sqrt{a_i} \sum_\mu \,  \sqrt{\omega_{\mu,i}} \,\, (\hat{b}_{\mu,i}+\hat{b}_{\mu,i}^\dagger) \,,
\end{equation}
where $a_i$ is a qubit-bath coupling strength.

In the limit of very weak coupling with the thermal reservoirs the reduced dynamics of the system is given by the following master equation \cite{Breuer2002},
\begin{equation} \label{eq:GME}
\frac{d \hat{\rho}_S}{dt}=-\frac{i}{\hbar}[\hat{H}_S,\hat{\rho}_S] + \sum_{i=1}^3 \mathcal{L}_i\{\hat{\rho}_S\}\,,
\end{equation}
where $\hat{\rho}_S$ is the system density matrix. The dissipation super-operators $\mathcal{L}_i\{\hat{\rho}_S\}$ are in the GKLS form 
\begin{multline} \label{eq:Li}
\mathcal{L}_i[\hat{\rho}_S(t)]= \\ \sum_{\omega>0} \gamma_i \, (1+n_\omega^i) \,  \, \left[ \hat{A}_\omega^i \,\hat{\rho}_S(t)\hat{A}_\omega^{i^\dagger} -\frac{1}{2}\hat{A}_\omega^{i^\dagger}\hat{A}_\omega^i \,\hat{\rho}_S(t) -\frac{1}{2}\hat{\rho}_S(t) \hat{A}_\omega^{i^\dagger}\hat{A}_\omega^i \right] \\ +  \gamma_i\,  n^i_{\omega} \,  \, \left[ \hat{A}_\omega^{i^\dagger} \, \hat{\rho}_S(t)\hat{A}_\omega^{i} -\frac{1}{2}\hat{A}_\omega^i \hat{A}_\omega^{i^\dagger} \, \hat{\rho}_S(t) -\frac{1}{2}\hat{\rho}_S(t) \hat{A}_\omega^i \hat{A}_\omega^{i^\dagger}\right] \,,
\end{multline}
with
\begin{equation} \label{eq:niw}
n_\omega^i= \frac{1}{\exp(\hbar\omega /k_BT_i)-1}
\end{equation}
the average thermal occupation of $R_i$ and $k_B$ the Boltzmann constant. The constant $\gamma_i$ describes the strength of the interaction between the system and $R_i$. Therefore $\gamma_i$ defines the lowest frequency scale in the model. The global jump operators $\hat{A}_\omega^i$ are obtained by using 
\begin{equation} \label{eq:fourierdecomposition}
e^{i\hat{H}_S t/\hbar}\hat{\sigma}_x^{\,i} \, e^{-i\hat{H}_S t/\hbar}=\sum_{\omega}\hat{A}_{\omega}^{i}  e^{-i\omega t} \,.
\end{equation}

Finally, the heat current $\dot{Q}_i(t)$ from the bath $R_i$ into the system is given by \citep{Breuer2002}
\begin{equation} \label{eq:GMEhc}
\dot{Q}_i(t)={\rm Tr }\{\hat{H}_S \mathcal{L}_i[\hat{\rho}_S(t)]\} \,.
\end{equation}
In the steady state, $\dot{Q}_i(t\to \infty)\equiv \dot{Q}_i$, the First and Second Law of Thermodynamics are expressed as
\begin{eqnarray}
\sum_{i=1}^3 \dot{Q}_i &=& 0 \,,  \label{eq:1st} \\
\dot{S}=-\sum_{i=1}^3 \frac{\dot{Q}_i}{T_i} &\geq& 0 \,,  \label{eq:2nd}  
\end{eqnarray}
where $\dot{S}$ is the total stationary entropy production. These relations stem directly from both the structure of the master equation \eqref{eq:GME} and the definition of the heat currents in Eq.~\eqref{eq:GMEhc} \cite{Spohn1978,Alicki1979}.

%--------------------------------------------------------------------------------------------------------------------------------

\subsection{Repeated interaction model} \label{local}

We consider now a model based on repeated interactions between each qubit $i$ and a reservoir $R_i$ consisting of a stream of auxiliary qubits described by the local Hamiltonian:
\begin{equation}
\hat{H}_{R_{i}}= \hbar B_i \, \hat{\tau}_z^{i},\label{eq:HRi}
\end{equation}
initially in equilibrium at temperature $T_i$, i.e. in the state $\exp(-\hat{H}_{R_{i}}/k_B T_i)/Z_i$ where $Z_i={\rm Tr}\exp(-\hat{H}_{R_{i}}/k_B T_i)$. Again, we assume $T_1<T_2<T_3$. The operators $\hat{\tau}_{x,y,z}^{i}$ are the corresponding Pauli operator for the auxiliary qubits. 
 The interaction between the system and each particle of the reservoir lasts for a short time $\tau$ and is described by the Hamiltonian
\begin{equation} \label{eq:HSR1}
\hat{H}_{SR_i}=\frac{\hbar \, g_i}{\sqrt \tau}\left( \hat{\sigma}_-^{i} \otimes \hat{\tau}_+^{i}+\hat{\sigma}_+^{i} \otimes \hat{\tau}_-^{i}\right) \,,
\end{equation}
where $\hat{\sigma}_\pm^i$ ($\hat{\tau}_\pm^i$) are the raising and lowering operators for the system (auxiliary) qubit $i$.
%
%with $\hat{b}^{i}$ and $\hat{b}^{i^\dagger}$ lowering and raising operators associated with a particle of the bath $R_i$. 

The model has been treated extensively in other publications \cite{Scarani2002,Ziman2002,Karevski2009,Giovannetti2012,Landi2014,Barra2015,StrasbergPRX2016,ManzanoPRE2016,Shammah2018,Seah2018,Ciccarello2017,PezzuttoNJP2016,pezzutto2018out,cusumano2018entropy,CampbellPRA2018,GrossQST2018,rodrigues2019thermodynamics,seah2019collisional,SeahPRE2019,strasberg2019repeated,ManPRA2019,ManQIP2019,ehrich2019micro} and here we only give a summary of the results. 
In the limit of weak coupling $g_i\tau\ll 1$ and taking $\tau\to 0$, the reduced dynamics of the system is described by the so called local master equation:
\begin{equation}
\frac{d \hat{\rho}_S}{dt}=-\frac{i}{\hbar}[\hat{H}_S,\hat{\rho}_S]+ \sum_{i}^{3}\mathcal{D}_i[\hat{\rho}_S],
\end{equation}
with $\mathcal{D}_i$ being the dissipator for the bath $R_i$
\begin{eqnarray}
\mathcal{D}_i[\hat{\rho}_S] &=&  \gamma_i \, (n_{2Bi}^i+1)\, \bigg( \hat{\sigma}_-^i \hat{\rho}_S \hat{\sigma}_+^i - \frac{1}{2}\hat{\sigma}_+^i \hat {\sigma}_-^i \hat{\rho}_S -\frac{1}{2} \hat{\rho}_S \hat{\sigma}_+^i \hat{\sigma}_-^i \,\bigg) \nonumber \\
&+& \gamma_i \, n_{2B_i}^i\, \bigg( \hat{\sigma}_+^i \hat{\rho} \hat{\sigma}_-^i - \frac{1}{2}\hat{\sigma}_-^i \hat{\sigma}_+^i \hat{\rho}_S -\frac{1}{2} \hat{\rho}_S \hat{\sigma}_-^i \hat{\sigma}_+^i \, \bigg). %\nonumber \\
\end{eqnarray}
 The parameter $\gamma_i=g_i^2$ is the rate resulting from the microscopic derivation of the local master equation using the repeated interaction model (see, for example, Ref.~\cite{De_Chiara_2018}).
 Unlike in the harmonic-bath model, the value of this constant is not necessarily smaller than the system inner frequencies. 

As for Sec.~\ref{global}, we are interested in the behaviour of the system at steady state. From the numerical expression of this stationary state $\hat{\rho}_S(t\rightarrow \infty)$, we can readily calculate the corresponding heat current flowing from each reservoir via,
\begin{equation}\label{eq:LMEheat}
\dot{Q}_i={\rm Tr}\{\hat{H}_{L_{i}} \mathcal{D}_i[\hat{\rho}_S(t\rightarrow \infty)]\} \,.
\end{equation}
The thermodynamic consistency of the model can only be ensured if one takes into account the extra work cost associated with the time dependence of the system-auxiliary qubit interaction as shown in Refs.~\cite{Barra2015,StrasbergPRX2016, De_Chiara_2018}. We emphasise that this work is produced/paid by the agent or field  (e.g. electric or magnetic) that is generating the interaction between the system and the auxiliary qubit during their collision. The expression for the work power at steady state reads
\begin{equation}\label{eq:work_LME}
\dot{W}={\rm Tr}\left\{\hat{H}_I\sum_{i}^{3} \mathcal{D}_i [\mathcal{\hat\rho}_S(t\rightarrow \infty)]\right\}=-\sum_{i}^{3}\dot{Q}_i \,.
\end{equation}
From the last equality it is evident that the First Law, for the system at steady state, is automatically verified. We are assuming the convention that negative work corresponds to work extracted/produced. The Second Law for this model is expressed as Eq. \eqref{eq:2nd}. Using Eqs. \eqref{eq:HL}, \eqref{eq:conserved} and \eqref{eq:LMEheat}, an additional relation for the currents in this model is found,
\begin{equation} \label{eq:modified_1st}
\sum_{i=1}^3\frac{\dot{Q}_i}{B_i}=0 \,.
\end{equation}
%

%--------------------------------------------------------------------------------------------------------------------------------
%
%--------------------------------------------------------------------------------------------------------------------------------

\section{Operating regimes}\label{sec:modes}

Having defined the thermodynamic quantities of the system we can characterise the corresponding operating regimes via the sign of $\dot{Q}_i$ and $\dot{W}$. Note that the relation $\dot{W}\equiv 0$ always holds in the harmonic-bath model. For various parameters of the 3-qubit system Hamiltonian, in a $\{\dot Q_1/\dot Q_3,\dot Q_2/\dot Q_3\}$ diagram, these regimes are limited by the boundaries
\begin{equation}\label{eq:zero}
\dot Q_1=0,\;\dot Q_2=0,
\end{equation}
the line $\dot{W}=0$ given by
\begin{equation}\label{zerowork}
\frac{\dot{Q}_2}{\dot{Q}_3}=\frac{\dot{Q}_1}{\dot{Q}_3}-1,
\end{equation}
and the line where the entropy production \eqref{eq:2nd} is zero,
\begin{equation}\label{zerodeltaS}
\frac{\dot{Q}_2}{\dot{Q}_3}=\frac{T_2\dot{Q}_1}{T_1\dot{Q}_3}-\frac{T_2}{T_3}.
\end{equation}

%%%%%%%%%%%%%%%%%%%%%%%%%%%%%%%%%
\begin{table*}[t]
	\centering
\begin{tabular}{|c|l|l|c|c|c|c|}
\hline
Label &Operating regime &Description& $\dot Q_1$ & $\dot Q_2$ & $\dot Q_3$ & $\dot W$ 
\\
\hline
I&$T_1$-Refrigerator&Dual-sink power-driven refrigerator &$>0$&$<0$&$<0$&$>0$
\\
\hline
II&$T_1$-Refrigerator&Dual-source power-driven refrigerator &$>0$& $>0$&$<0$&$>0$
\\
\hline
III&$T_1$-Refrigerator& Power- and heat-driven refrigerator &$>0$& $<0$ &$>0$&$>0$
\\
\hline
IV&$T_1$-Refrigerator${^*}$& Absorption refrigerator &$>0$& $<0$&$>0$&$\le 0$
\\
%\hline
\noalign{\hrule height 1.5pt}
V&$T_1$-Heater & $T_2$-power-driven refrigerator&$<0$&$>0$&$<0$&$>0$
\\
\hline
VI&$T_1$-Heater${^*}$& $T_2$-heat-driven refrigerator with work production &$<0$& $>0$&$<0$&$\le 0$
\\
\hline
VII&$T_1$-Heater& Dual-source accelerator&$<0$&$>0$&$>0$&$>0$
\\
\hline
VIII&$T_1$-Heater${^*}$& Dual-source engine &$<0$& $>0$&$>0$&$\le 0$
\\
\hline
IX&$T_1$-Heater & Dual-sink accelerator &$<0$& $<0$&$>0$&$>0$
\\
\hline
X&$T_1$-Heater${^*}$ &Dual-sink thermal engine&$<0$&$<0$&$>0$&$\le 0$
\\
\hline
\end{tabular}

\caption{Table of the different operating regimes of the three-qubit machine. The first column shows the labels used in the text and the in subsequent plots; the second column describes the operating regime by taking as a reference the currents exchanged with $R_1$ while the third column further describes the operation of the setup as a thermal machine (notice that this might not be unique); the remaining columns show the signs of the heat currents and of the external work power. The asterisk indicates the operating regimes accessible in the harmonic-baths model with the requirement $\dot W=0$. All operating regimes can be realised in the repeated interactions model.} \label{fig:operatingmodes}
\end{table*}
%%%%%%%%%%%%%%%%%%%%%%%%%%%%%%%%% 

The possible operating regimes are summarised in Table~\ref{fig:operatingmodes} depending on the signs of $\dot{Q}_i$ and $\dot{W}$. As we will see, all ten modes can be realised in the repeated interaction model. For the harmonic-bath model, since  $\dot{W}=0$, only four modes, indicated with an asterisk in Table~\ref{fig:operatingmodes}, are possible. We have classified the operating regimes into two main categories: the refrigerators related to an extraction of energy from $R_1$ and the heaters that lead to an injection of energy into $R_1$. In particular, regime IV corresponds to absorption refrigerators. Such functioning allows for the cooling of the cold bath without the input of external work. This cooling mode is driven by the heat coming from $R_3$. 

For an absorption refrigerator, the performance can be assessed by the amount of heat extracted from $R_1$ compared to that inputted from $R_3$. This means that, in absence of work extracted, the coefficient of performance (COP) for the machine is the ratio ${\rm COP}=\dot{Q}_1/\dot{Q}_3$. We see that the maximum possible value of $\dot{Q}_1/\dot{Q}_3$ is found where the lines $\dot{S}=0$ and $\dot{W}=0$ intersect, corresponding to the maximum COP 
\begin{equation}
\label{eq:copcarnot}
{\rm COP}_{\rm max}=\dfrac{T_1(T_3-T_2)}{T_3(T_2-T_1)}
\end{equation}
which only depends on the temperature ratios. To assess the viability of an absorption refrigerator considering the different environment models, we begin with a random search of the parameters space.

\subsection{Harmonic-bath model}

\begin{figure}[b]
	\centering
	\includegraphics[width=0.85\columnwidth]{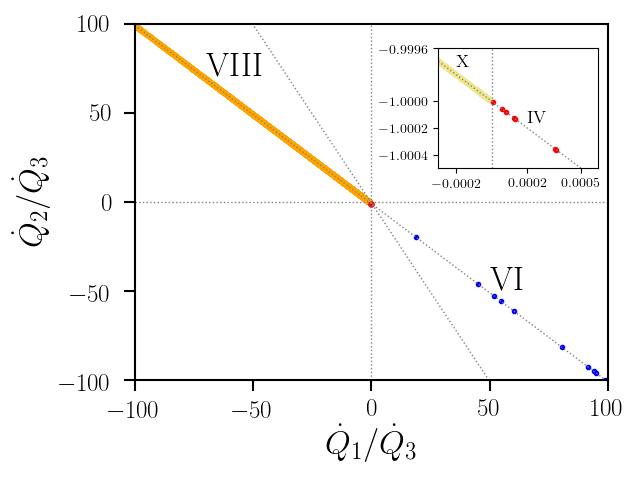}
	\caption{Scatter plots of the heat ratios $\frac{\dot{Q}_1}{\dot{Q}_3}$ and $\frac{\dot{Q}_2}{\dot{Q}_3}$ for $50,000$ random parameter choices of the local magnetic fields $B_i$ in the interval $(0,1)$. The red, blue, orange and khaki dots corresponds to the operating regimes IV, VI, VIII and X. 
	%The vertical dashed line is located at $\frac{\dot{Q}_1}{\dot{Q}_3}={\rm COP}_{max}$ (Eq. \eqref{eq:copcarnot}). 
	%This set the boundary between the refrigerators and heat transformers. 
	Equations  \eqref{eq:zero}, \eqref{zerowork} and \eqref{zerodeltaS} are  represented by using dotted lines. The other parameters are $\gamma_1=8.71\cdot 10^{-7}$, $\gamma_2=5.76\cdot 10^{-7}$, $\gamma_3=7.56\cdot 10^{-7}$, $\Delta_{12}=7.93\cdot 10^{-4}$, $\Delta_{13}=9.67\cdot 10^{-4}$, $\Delta_{23}=1.69\cdot 10^{-4}$, $J_{12}=5.49\cdot 10^{-4}$, $J_{13}=2.960\cdot 10^{-4}$, $J_{23}=4.963\cdot 10^{-4}$, $T_1=1$, $T_2=2$, $T_3=3$.} \label{fig:GME_operation_modes}
\end{figure}

When considering the harmonic-bath model, there exist four possible operating regimes for the three-qubit machine. Namely, the device can act either as an absorption refrigerator IV or as a heater. There are three different regimes corresponding to the heater category VI, VIII and X. In Table~\ref{fig:operatingmodes} we show the sign of the currents $\dot{Q}_i$ corresponding to each one of these regimes of operation. In Fig.~\ref{fig:GME_operation_modes} we show the different regimes of three-qubit machine obtained with random local magnetic fields $B_i$ sampled with uniform distribution in the interval $[0,1]$. Values of $B_i$ of the order of $\gamma$ were discarded. The consistency with the First Law is reflected in the fact that all the regimes of operation lie on the line defined by Eq. \eqref{zerowork}. These results indicate that the most likely operations are either VIII or X. However, the absorption refrigerators and the heater VI are found for more specific parameter regimes. Concretely, the results corresponding to absorption refrigerators are very close to the origin. Thus, the corresponding performance turns out to be very small. 
%This cooling operation can be understood as the interplay between two main mechanisms. One of them is associated with the unavoidable heat leaks from the baths $R_3$ and $R_2$ into $R_1$. The other is related to three-bath processes that make it possible to cool down $R_1$. Cooling conditions are only found when this second mechanism overcomes the heat leaks. This occurs typically for small internal couplings ($J_{ij}/B_i,\Delta_{ij}/B_i \ll 1$) while bath temperatures should be of comparable magnitude. Under these conditions the heat leaks are minimized at the expense of small internal couplings and, as a consequence, small currents. Therefore, although cooling is achievable by using two-body interactions in harmonic-bath models, three-body couplings are more convenient to obtain large currents and performance, see e.g. \cite{LindenPRL2010}.
This cooling operation can be understood as the interplay between two main mechanisms. One of them is associated with the unavoidable heat leaks from the baths $R_3$ and $R_2$ into $R_1$. The other is related to three-bath processes that make it possible to cool down $R_1$. The currents associated with both mechanisms are proportional to transition frequencies between the eigenstates of $\hat{H}_S$. Therefore, the cooling mechanism only overcomes heat leaks for very specific values of the parameters, typically for small internal couplings ($J_{ij}/B_i,\Delta_{ij}/B_i\ll 1$) while bath temperatures should be of comparable magnitude. Under these conditions the heat leaks are minimized at the expense of small internal couplings and, as a consequence, small currents. For this reason two-body interactions are less convenient to obtain large currents and performance than three-body couplings \cite{LindenPRL2010}. In the latter case, the currents associated with the heat leaks are proportional to the coupling strength between qubits \cite{CorreaPRE2015}. As a result, the cooling operation is found for a larger set of parameters, occurring in addition with larger currents and performance. When the internal couplings are of the order of the qubit frequencies, heat leaks will dominate in both models and cooling conditions cannot be found.

%
%\begin{table}[H]
%\begin{tabular}{|l|c|c|c|}
%\hline
%Operation & $\dot Q_1$ & $\dot Q_2$ & $\dot Q_3$ 
%\\
%\hline
%Dual-sink &$<0$&$<0$&$>0$
%\\
%\hline
%$T_2$-refrigerator &$<0$&$>0$&$>0$
%\\
%\hline
%Absorption refrigerator &$>0$& $<0$&$>0$
%\\
%\hline
%Heat transformer &$<0$& $>0$&$<0$
%\\
%\hline
%\end{tabular} 
%\caption{Table of the different operation modes corresponding to the harmonic-bath model.} \label{tab:modesGME}
%\end{table}
%%%%%%%%%%%%%%%%%%%%%%%%%%%%%%%%%
%%%%%%%%%%%%%%%%%%%%%%%%%%%%%%%%%%

\subsection{Repeated interaction model}

The repeated interaction model should include the work cost associated with the microscopic collisions to ensure thermodynamic consistency. This additional energy current is responsible for the emergence of ten different regimes of operation. We begin by fixing $J_{i,j},\Delta_{i,j}$ to some randomly chosen values in the range $[0,1]$; we observe that the results do not qualitatively depend on this initial choice. We fix the temperatures with $T_1=1~,~T_2=2~,~T_3=3$. A large set of the remaining parameters $ \{B_i,\gamma_i\} $ are then randomly generated with the restriction $0\leq B_i \leq 5$ and $0\leq \gamma_i \leq 1$. The resulting steady states and currents are calculated. 
%%%%%%%%%%%%%%%%%%%%%%%%%%%%%%%%%
\begin{figure}[t]
	\centering
	\includegraphics[width=0.9\columnwidth]{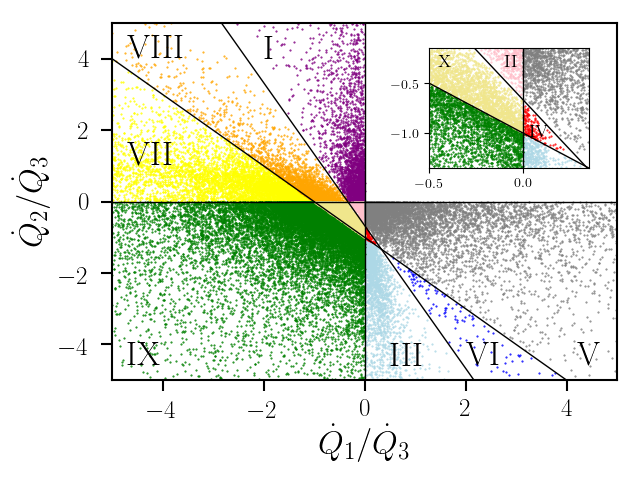}
	\caption{Main panel: A scatter plot of the heat ratios ${\dot{Q}_1}/{\dot{Q}_3}$ and ${\dot{Q}_2}/{\dot{Q}_3}$ for $50,000$ random parameter choices of the local magnetic fields $B_i$ and dissipation rates $\gamma_i$. The 
%	red, pink, purple, blue, light blue, grey, yellow, orange, green and light green 
purple, pink, light blue, red, grey, blue, yellow, orange, green and khaki	dots correspond respectively to the operating regimes I, II, III, IV, V, VI, VII, VIII, IX and X. The lines correspond to Eqs.~\eqref{eq:zero}, \eqref{zerowork} and \eqref{zerodeltaS}. The other parameters are $J_{12}=9.81\cdot 10^{-1}, J_{13}=7.75\cdot 10^{-1}, J_{23}=7.57\cdot 10^{-1}$,
	$\Delta_{12}=1.24\cdot 10^{-1}, \Delta_{13}=2.56 \cdot 10^{-1}, \Delta_{23}=6.11\cdot 10^{-1}$. In the inset we show the same diagram but restricted to the absorption refrigerators.}	\label{fig:random}
\end{figure}
%%%%%%%%%%%%%%%%%%%%%%%%%%%%%%%%% 

The heat current ratios ${\dot{Q}_1}/{\dot{Q}_3}$ and ${\dot{Q}_2}/{\dot{Q}_3}$  are presented in the scatter plot of Fig. \ref{fig:random}. We see that there is a far richer selection of possible operating regimes compared to the harmonic-bath model. They are summarised in the Table \ref{fig:operatingmodes}. 

We note the top right quadrant is completely empty, this is due to the fact that having both ${\dot{Q}_1}/{\dot{Q}_3}$ and ${\dot{Q}_2}/{\dot{Q}_3}$ positive indicates that all heat currents flow in the same direction. This is, however, not possible because of the conservation of magnetisation of the system: as a consequence of Eq. \eqref{eq:modified_1st}, these currents cannot have all the same sign.

Focusing on our intended target of designing an absorption refrigerator we note that the requirements  $\{\dot{Q}_1>0,\dot{Q}_2<0,\dot{Q}_3>0,\dot{W} \leq 0\}$ are fulfilled within the small triangle in the bottom right quadrant of Fig. \ref{fig:random}. This area is constrained by $\dot S=0$, $\dot{W}=0$ and ${\dot{Q}_1}/{\dot{Q}_3}=0$, (label IV) and highlighted in the inset of Fig.~\ref{fig:random}. All the points above the line $\dot W = 0$ correspond to negative work power, i.e. besides refrigeration of $R_1$ there is some extra work produced. In this case the COP defined above is smaller than the maximum value. Therefore a fairer definition of COP in this case might be:
\begin{equation}
{\rm COP}_W =\frac{ \dot{Q}_1}{\dot{Q}_3+\dot W}.
\end{equation}

As we have shown in this section, it is possible to construct an absorption refrigerator with three qubits and two-body interactions also when the total system is described by the local master equation. In the next section, we will elucidate how such 3-qubit machines can be understood as coupled 2-reservoir machines. 

%--------------------------------------------------------------------------------------------------------------------------------

\subsubsection{Decomposition in two-reservoir devices} \label{sec:decomposition}

In this section we show how the three-qubit machine in the context of the repeated collision model can be decomposed into three coupled two-reservoir devices. Then, we will analyze the absorption refrigerator as a composite machine, in which one of the devices operates as an engine between the work reservoir at $T_3$ and the hot reservoir at $T_2$ whereas a second device operates as a refrigerator between the hot and cold ($T_1$) reservoirs. The work produced from the engine is used to power the refrigerator. This insight allows us to restrict considerably the choice of magnetic fields necessary to design one of the operating regimes described in the previous section. 

The internal functioning of the three-qubit setup can be further clarified calculating the magnetic current moving throughout the system \cite{PhysRevE.94.042122}:
\begin{equation}
\label{eq:continuity}
\frac{d \langle \sigma_z^i \rangle}{dt}= \dot q_i +\sum_{j\neq i} C_{j,i},
\quad i=1,2,3
\end{equation}
where $C_{j,i} = 2 J_{ji} \langle \sigma_x^j\sigma_y^i-\sigma_x^i\sigma_y^j \rangle=-C_{i,j}$ is the magnetic current from qubit $j$ to $i$ and $\dot q_i = \gamma_i(1+2 \bar{n}_i) \Big( \langle \sigma_z \rangle_{b_i}-  \langle \sigma_z \rangle_{i}\Big)$ is the magnetic current between qubit $i$ and its corresponding bath. Summing Eq.~\eqref{eq:continuity} over $i=1,2,3$ we obtain:
\begin{equation}
\sum_{i=1,2,3} \frac{d \langle \sigma_z^i \rangle}{dt}=\sum_{i=1,2,3}\dot q_i,
\end{equation}
implying that the variation of the total magnetisation of the system is given by the magnetic flux from the three reservoirs. 
Furthermore when the system is in a steady state, $\sum_i{d \langle \sigma_z^i \rangle}/{dt}=0$, we obtain the relation
\begin{equation}\label{magcurrentsum}
\sum_{i=1,2,3} \dot q_i=0\,.
\end{equation}
This condition is equivalent to Eq. \eqref{eq:modified_1st} since the heat currents are proportional to the magnetic currents,
\begin{equation}\label{eq:relation}
\dot{Q}_i= B_i \dot q_i.
\end{equation}
In the case in which all the magnetic fields are equal: $B_i=B$, the condition $\sum_i \dot q_i=0$ implies $\sum_i \dot Q_i=0$ which means that the external work power $\dot W$ is zero, (see Eq. \eqref{eq:work_LME} and Ref. \cite{De_Chiara_2018}). This correspondence between the balance of magnetic currents and that of heat currents does not hold in the case of unequal fields. 
 
\begin{figure}[t]
\begin{center}
\includegraphics[width=0.9\columnwidth]{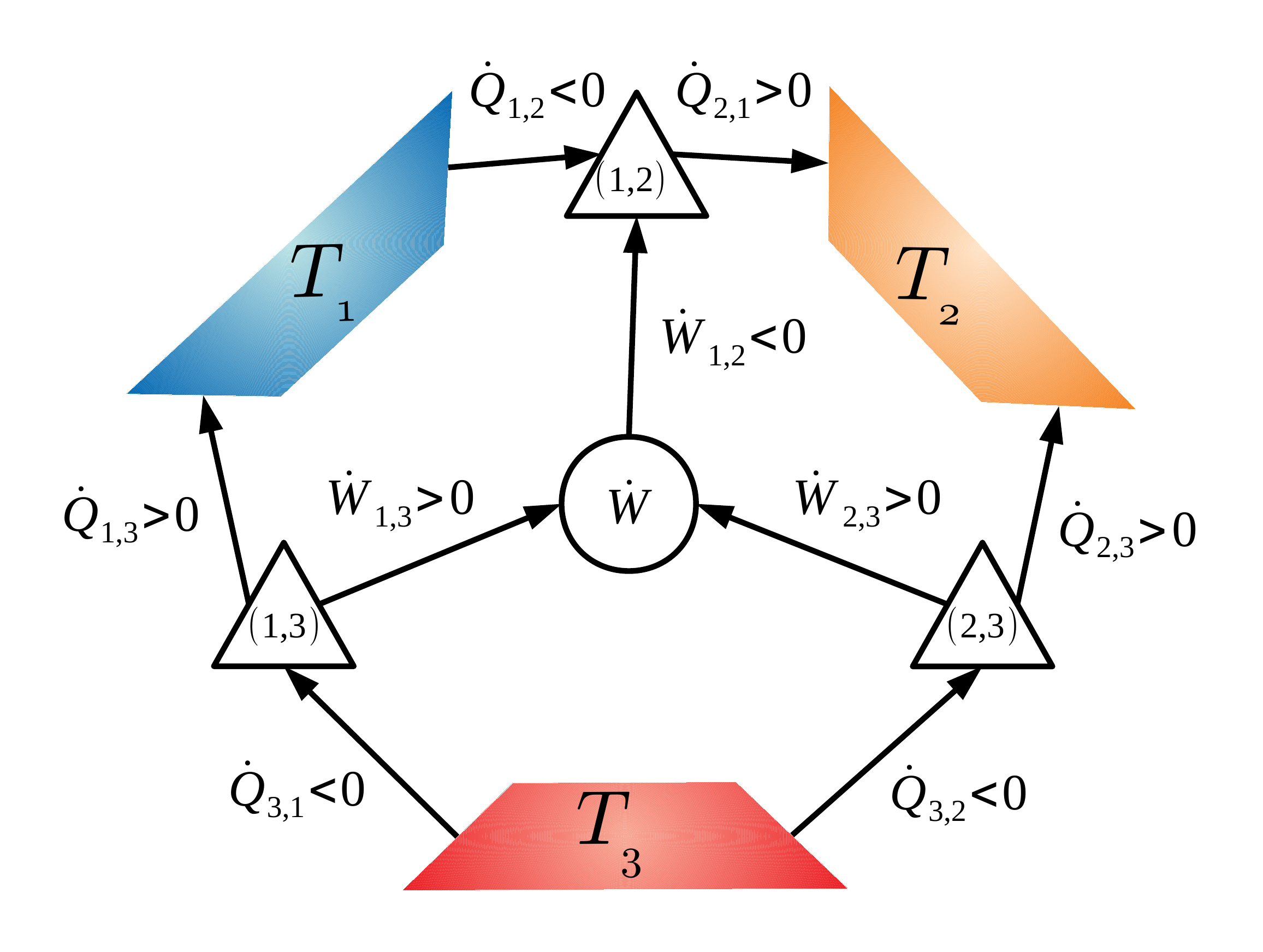}
\caption{Schematic representation of the three-qubit absorption refrigerator decomposed as the sum of three two-reservoir devices depicted as triangles. We show a particular case corresponding to the absorption refrigerator mode IV for which the pairs $(2,3)$ and $(1,3)$ act as engines, while $(1,2)$ operates as a refrigerator.}
\label{fig:submachines}
\end{center}
\end{figure}

The continuity equation \eqref{eq:continuity} allows us to see how the three-qubit system can be thought of as three two-reservoir devices. A schematic view of this construction is shown in Fig. \ref{fig:submachines}. 
%%%%%%%%%%%%%%%%%%%%
%%%%%%%%%%%%%%%%%%%%
 Each pair of qubits has an associated magnetic current $C_{i,j}$ with a specific sign. This magnetisation current is associated with a corresponding energy current but due to the difference in the local magnetic fields that the two qubits may have, the energy currents that flow out of one qubit and into another are not equal $ B_i C_{i,j} \neq - B_j C_{j,i}$. From its definition, $C_{j,i}$ can be interpreted as the rate of excitations exchange between the two qubits. When the local magnetic fields are unequal, the energy released by one qubit flipping is different from that absorbed by the other qubit making the reverse flip. This difference can be interpreted as a work power associated with a 2-qubit device labelled $(i,j)$:
 \begin{equation}
\dot{W}_{i,j}=(B_i-B_j) C_{i,j} = -\dot Q_{i, j}-\dot Q_{j, i}.
\end{equation}
where we have defined the heat currents $\dot Q_{i, j}=-B_iC_{i,j}$ from qubit $j$ to qubit $i$. In Appendix~\ref{sec:corre}, we show that large currents $C_{i,j}$ are necessarily accompanied by large quantum correlations measured, for instance, by the quantum mutual information. 

Notice that the total work power produced by the three-qubit system is the sum of the three contributions from the three devices:
 \begin{equation}
 \label{eq:totalWsubmachines}
-\dot{W}=\dot W_{1,2}+ \dot W_{1,3}+\dot W_{2,3}.
\end{equation}

Each of the three 2-qubit devices seems to operate similarly to a system of two qubits coupled by the XXZ interaction, subject to the local magnetic fields $B_i$ and $B_j$, respectively, and coupled to the reservoirs at temperatures $T_i$ and $T_j$. Another interesting feature is that each 2-qubit device has the same expression for the efficiency as the one corresponding to a single Otto engine or refrigerator. Such machine would consist of a qubit operating with the same magnetic fields $B_i$ and $B_j$. This originates from the mathematical structure of the continuity equation that implies that the ratios of heat currents and of heat current and work power only depend on the local magnetic fields applied to the qubits $i$ and $j$. For example, if the $(i,j)$ device operates as a thermal engine, then its efficiency is: 
\begin{equation}
\eta_{i,j} = \frac{|\dot{W}_{i,j}|}{\dot Q_{j, i}} = 1-\frac{B_i}{B_j}.
\end{equation}
Conversely, if it operates as a refrigerator, its coefficient of performance is:
\begin{equation}
{\rm COP}_{i,j} = \frac{|\dot Q_{i, j}|}{\dot{W}_{i,j}} = \frac{B_i}{B_j-B_i}.
\end{equation}
Remarkably, there is an important caveat: the three devices are not independent since they are connected to each other as they exchange the work contributions $\dot W_{i,j}$ which are constrained by Eq.~\eqref{eq:totalWsubmachines}. Thus, even if individually they operate similarly to Otto machines, such components may violate the Carnot bounds for some parameters. However, the whole system obeys the Carnot limit. Moreover, it is not possible, based on the three magnetic fields $\{B_1,B_2,B_3\}$ and the three temperatures $\{T_1,T_2,T_3\}$, to predict the functioning of the collective machine without finding the actual steady state of the three-qubit system.

%%%%%%%%%%%%%%%%%%%%%%%%%%%%%%%%%
\begin{figure}[t!]
	\centering
	\includegraphics[width=0.9\columnwidth]{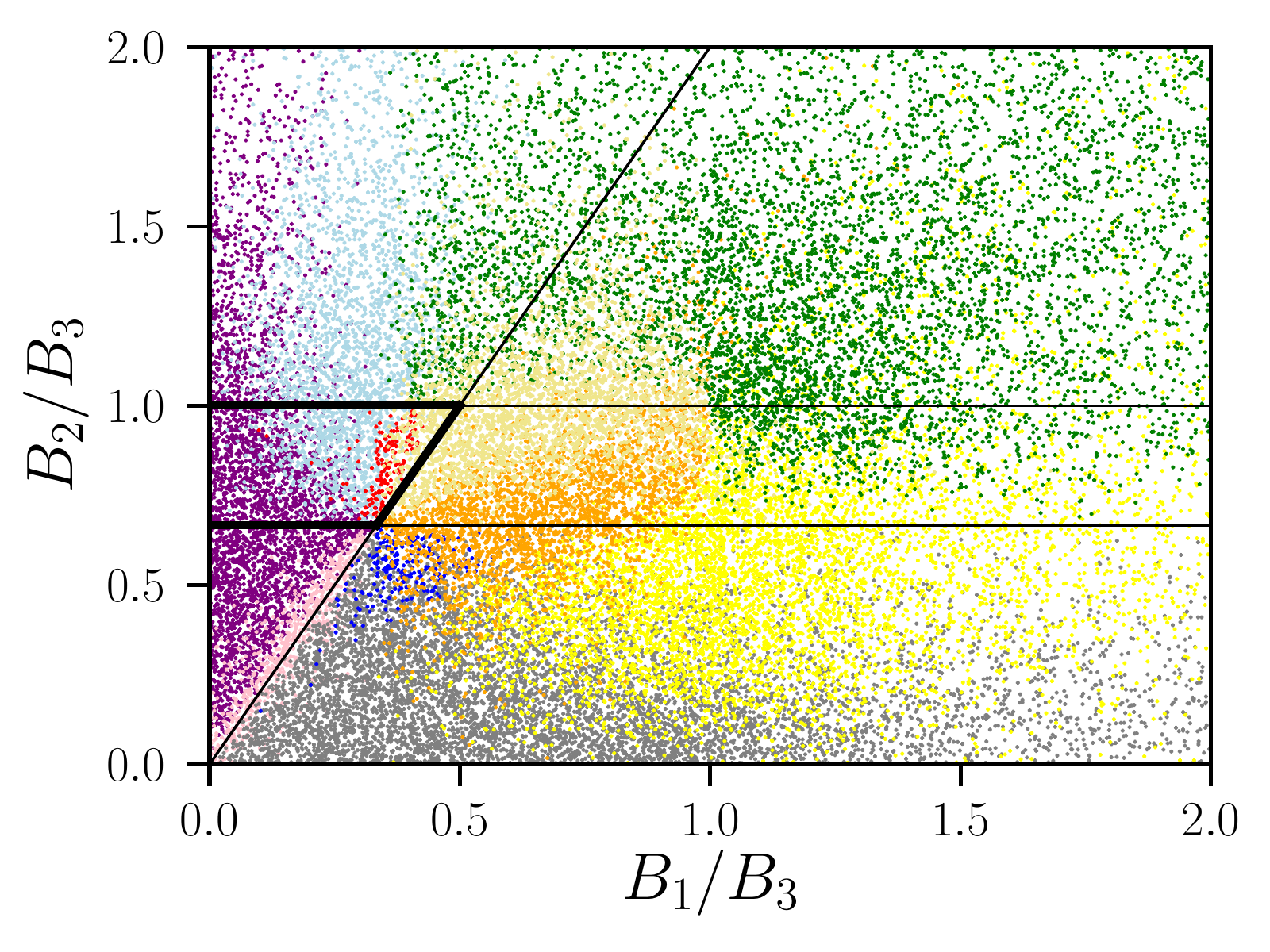}
	\caption{Scatter plot of the same $50,000$ points as Fig. \ref{fig:random} as a function of the field ratios ${B_1}/{B_3}$ and ${B_2}/{B_3}$. The black lines correspond to the operating regimes from Eq.~\eqref{FieldRestricsions} and the thick segments delimit the region where the absorption refrigerator (red points) can exclusively been found.	
	 }
	\label{fig:fields}
\end{figure}
%%%%%%%%%%%%%%%%%%%%%%%%%%%%%%%%%

Nevertheless this construction is useful to {\it exclude} the region of parameters in which the three-qubit machine cannot function. For example, if we consider the absorption refrigerator again, then we notice that for this to operate it must include internally an engine drawing heat from the warmest reservoir $T_3$ and producing some work which is then used to feed a refrigerator extracting heat from the coldest reservoir $T_1$. In the range of parameters corresponding to Fig.~\ref{fig:fields}, we have checked numerically that the pair $(2,3)$ act as an engine and the pair $(2,1)$ is a refrigerator. Besides, both satisfy the Carnot limit. Assuming $T_i<T_j$, an Otto machine works as a refrigerator for~\cite{PhysRevA.98.042102}
\begin{equation}
\label{eq:conditionrefrigerator}
\frac{B_i}{B_j}<\frac{T_i}{T_j},
\end{equation}
 as an engine for 
 \begin{equation}
 \label{eq:conditionengine}
\frac{T_i}{T_j}<\frac{B_i}{B_j}<1
\end{equation}
 and as an accelerator for $B_i/B_j>1$. Then, considering \eqref{eq:conditionrefrigerator} for the device $(2,3)$ and  \eqref{eq:conditionengine} for $(2,1)$, see Fig. \ref{fig:submachines}, lead to the following inequalities,
\begin{eqnarray}
\dfrac{T_2}{T_1} \dfrac{B_1}{B_3}<&\dfrac{B_2}{B_3}& \nonumber
\\
\frac{T_2}{T_3}<&\dfrac{B_2}{B_3}&<1,
\label{FieldRestricsions}
\end{eqnarray}
which delimit a trapezoid in a $(B_1/B_3,B_2/B_3)$ diagram (see Fig.~\ref{fig:fields}). We have also checked numerically that  the operating regime corresponding to the other pair $(1,3)$ can either be a refrigerator or an engine. However, this does not result in an additional restriction since such device may violate the Carnot bound. 

To verify these conditions, we plot again the same points of Fig.~\ref{fig:random} as a function of their magnetic ratios  $B_1/B_3$ and  $B_2/B_3$. The resulting scattering plot is in Fig.~\ref{fig:fields} which includes the straight lines corresponding to the conditions \eqref{FieldRestricsions}. In contrast to Fig.~\ref{fig:random} however, different operating regimes overlap because of the collective effects in the three two-qubit device model.

We observe that the red points, corresponding to the absorption refrigerators, are only confined to the trapezoid delimited by conditions \eqref{FieldRestricsions}. We note that this region is not solely the domain of the absorption refrigerator but contains three other regimes within it (see Table~\ref{fig:operatingmodes}): I, III  and X. Notice that in both regions I and III, the three-qubit system dissipates mechanical work rather than producing it as in the absorption refrigerator area IV. These functionings are not incompatible with conditions \eqref{FieldRestricsions}. In fact, because of the couplings between devices, it may happen that the work produced by the (2,3) device is not enough to power the refrigerator (1,2) thus turning the setup into one of the other modes. It is important to stress that the magnetic field conditions \eqref{FieldRestricsions} are necessary but not sufficient for designing an absorption refrigerator.

%--------------------------------------------------------------------------------------------------------------------------------
 
\subsubsection{Recycling the collisional work cost}

%%%%%%%%%%%%%%%%%%%%%%%%%%%%%%%%%
\begin{figure}[h!]
	\centering
	\includegraphics[width=0.9\columnwidth]{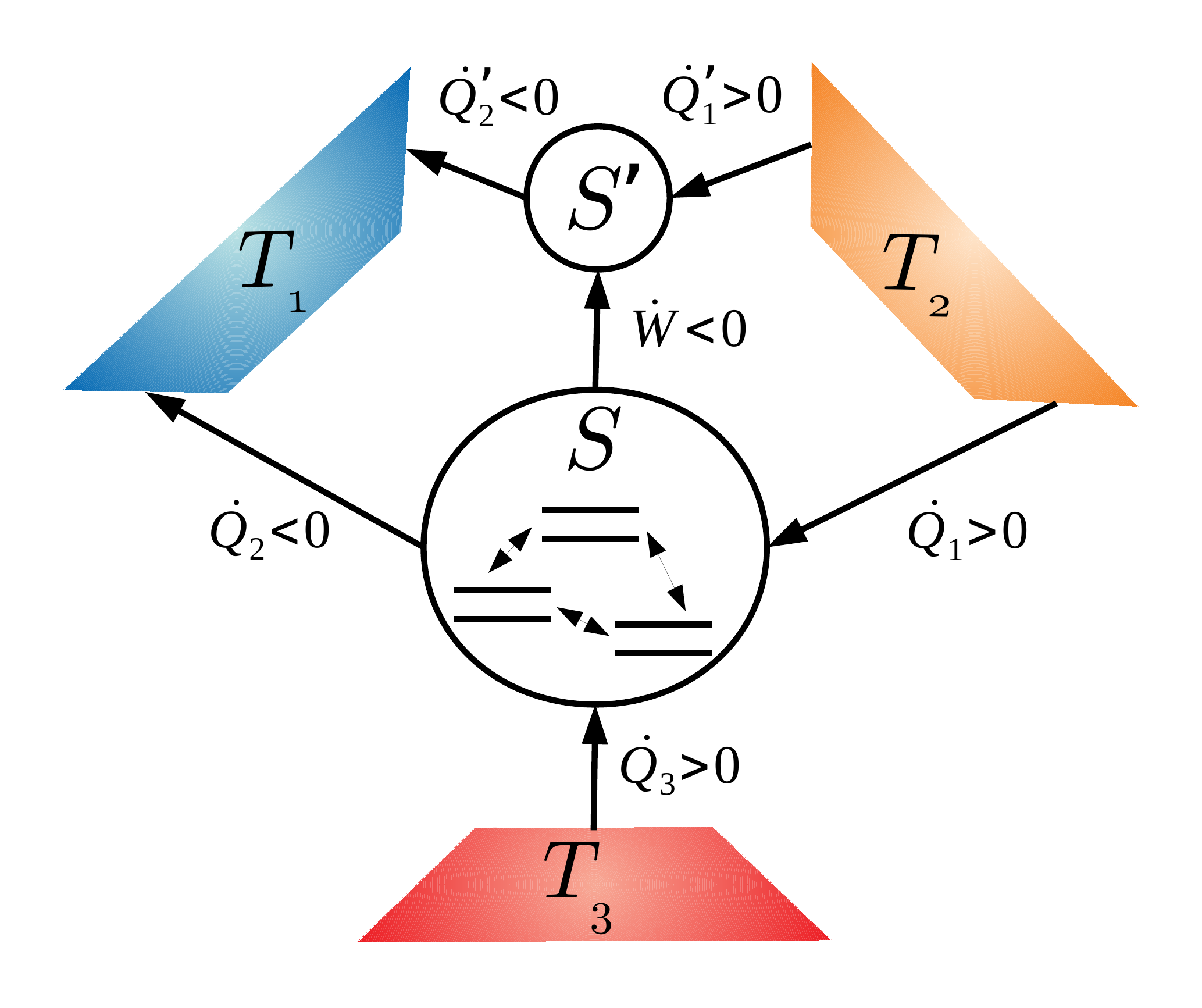}
	\caption{Scheme of composite machine recycling the work cost. $S$ is the original system of three qubits in the repeated interactions model and $S'$ is an Otto chiller . Both systems are coupled to the same baths $R_i$ and operate as refrigerators, i.e. $\{\dot{Q}_1>0,\dot{Q}_2<0,\dot{Q}_3>0,\dot{W}<0\}$ and $\{\dot{Q}'_1>0,\dot{Q}'_2<0,-\dot{W}>0\}$. $S'$ harness the work produced by the three-qubit device to boost the performance of the composite machine. }
	\label{fig:recycling}
\end{figure}
%%%%%%%%%%%%%%%%%%%%%%%%%%%%%%%%%

\begin{figure}[t]
\begin{center}
\includegraphics[width=0.9\columnwidth]{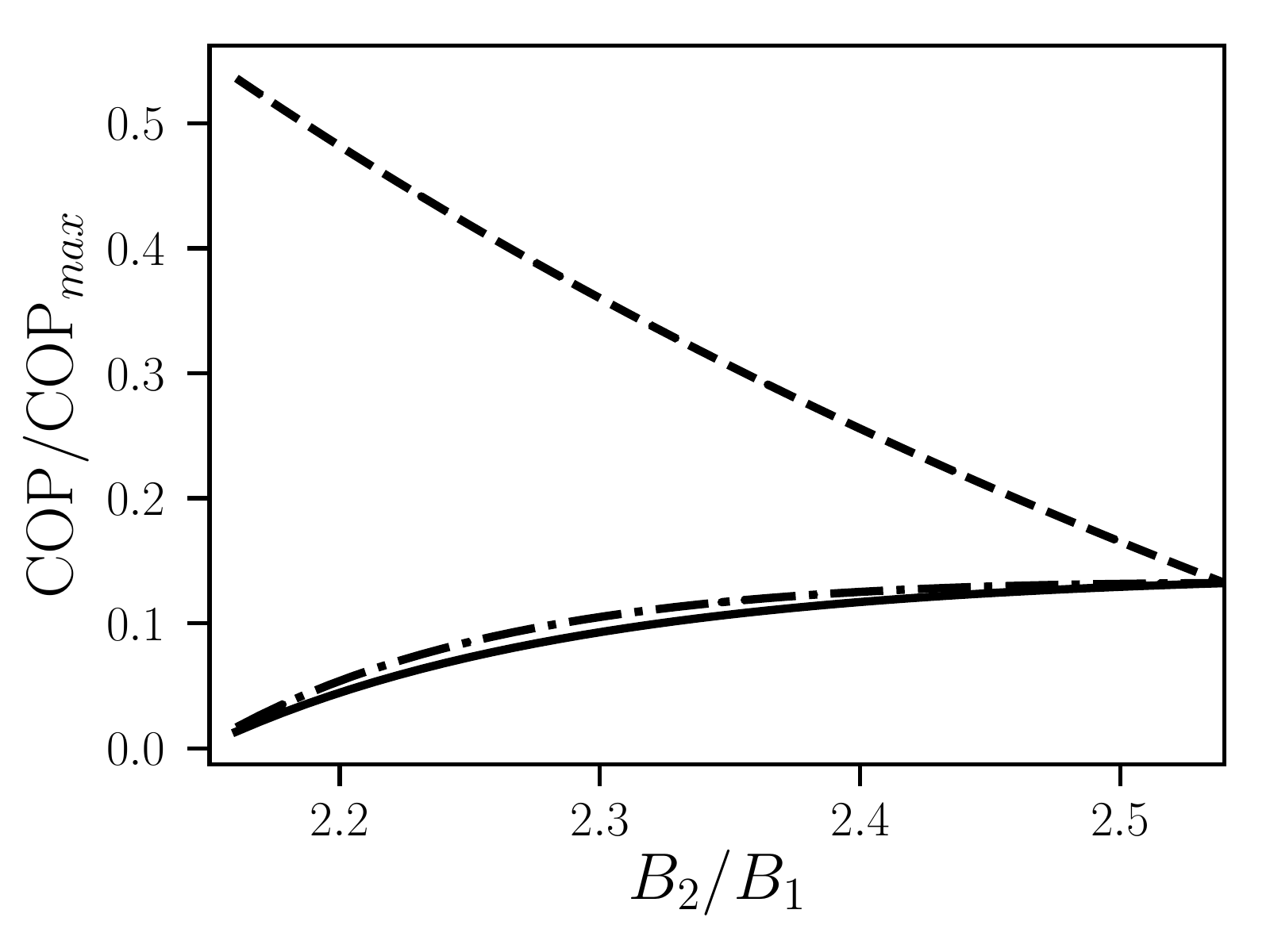}
\caption{The coefficients of performance ${\dot{Q}_1}/{\dot{Q}_3}$ (solid line), $\frac{\dot{Q}_1}{\dot{Q}_3+\dot{W}}$ (dotted-dashed line) and $\frac{\dot{Q}_1+\dot{Q}_1'}{\dot{Q}_3}$ (dashed line) in units of ${\rm COP}_{\rm max}$ (see \eqref{eq:copcarnot}) as functions of $B_2$. We have chosen $B_1=1.31$, $B_3=3.57$, $\gamma_1=6.45\cdot 10^{-1}$, $\gamma_2=7.80\cdot 10^{-1}$, $\gamma_3=9.34\cdot 10^{-1}$. Besides, we only show the values of $B_2$ such that both the original three-qubit system and the composite device operates as absorption refrigerators. The maximum value of $B_2$ corresponds to $\dot{W}=0$. The remaining parameters are the same as in Fig. \ref{fig:random}. }
\label{fig:performance_boost}
\end{center}
\end{figure}

The performance of the refrigeration process can be boosted if the extra work produced is recycled to feed a (classical or quantum) refrigerator $S'$ operating between $T_1$ and $T_2$ and requiring external work. Let us consider the scheme shown in Fig. \ref{fig:recycling} where $S$ is the original three-qubit refrigerator described by the repeated interaction model which produces some work. Let us also assume that the second refrigerator  $S'$ is an Otto fridge using a single qubit working with the magnetic field values $B_1$ and $B_2$. Its coefficient of performance is $B_1/(B_2-B_1)$ when operating in the adiabatic regime \cite{Tova2000}. If we combine the quantum 3-qubit machine with this Otto refrigerator, then  the whole performance will only depend on the field ratios:
\begin{equation} \label{eq:otto}
{\rm COP}_{\rm Otto}=\frac{\dot{Q}_1+\dot{Q}_1'}{\dot{Q}_3}=\frac{B_1(B_3-B_2)}{B_3(B_1-B_2)}
\end{equation}
corresponding to that of an Otto machine operating with three magnetic fields in the adiabatic regime. This last equation follows directly from Eq. \eqref{eq:modified_1st}. Interestingly, this performance is always greater than the performance of the original system. In Fig. \ref{fig:performance_boost} we illustrate this performance boosting.

%%%%%%%%%%%%%%%%%%%%
%%%%%%%%%%%%%%%%%%%%

%--------------------------------------------------------------------------------------------------------------------------------
%
%--------------------------------------------------------------------------------------------------------------------------------

\section{Heat Valve}\label{sec:valve}

Recently there has been a great deal of work investigating the possibility of building systems that can control the flow of heat currents much like what can be done with electrical currents. Schemes for quantum thermal transistors, which modulate a heat current by the application of a small auxiliary current and heat valves, which can reverse the direction of a current have been put forward  \cite{PhysRevLett.116.200601,PhysRevE.98.022118,GuoPRE2019, mandarino2019thermal} and experimentally realised~\cite{ronzani2018tunable}.

In addition, heat current amplification, where the magnitude of a current can be increased by application of a small auxiliary current, and current stabilisation, where the currents are unaffected by parameter changes, have been observed \cite{PhysRevLett.116.200601,PhysRevE.99.032114,GuoPRE2019,PhysRevE.98.022118,PhysRevA.97.052112}. 

\begin{figure}[t]
	\centering
	\includegraphics[width=0.9\columnwidth]{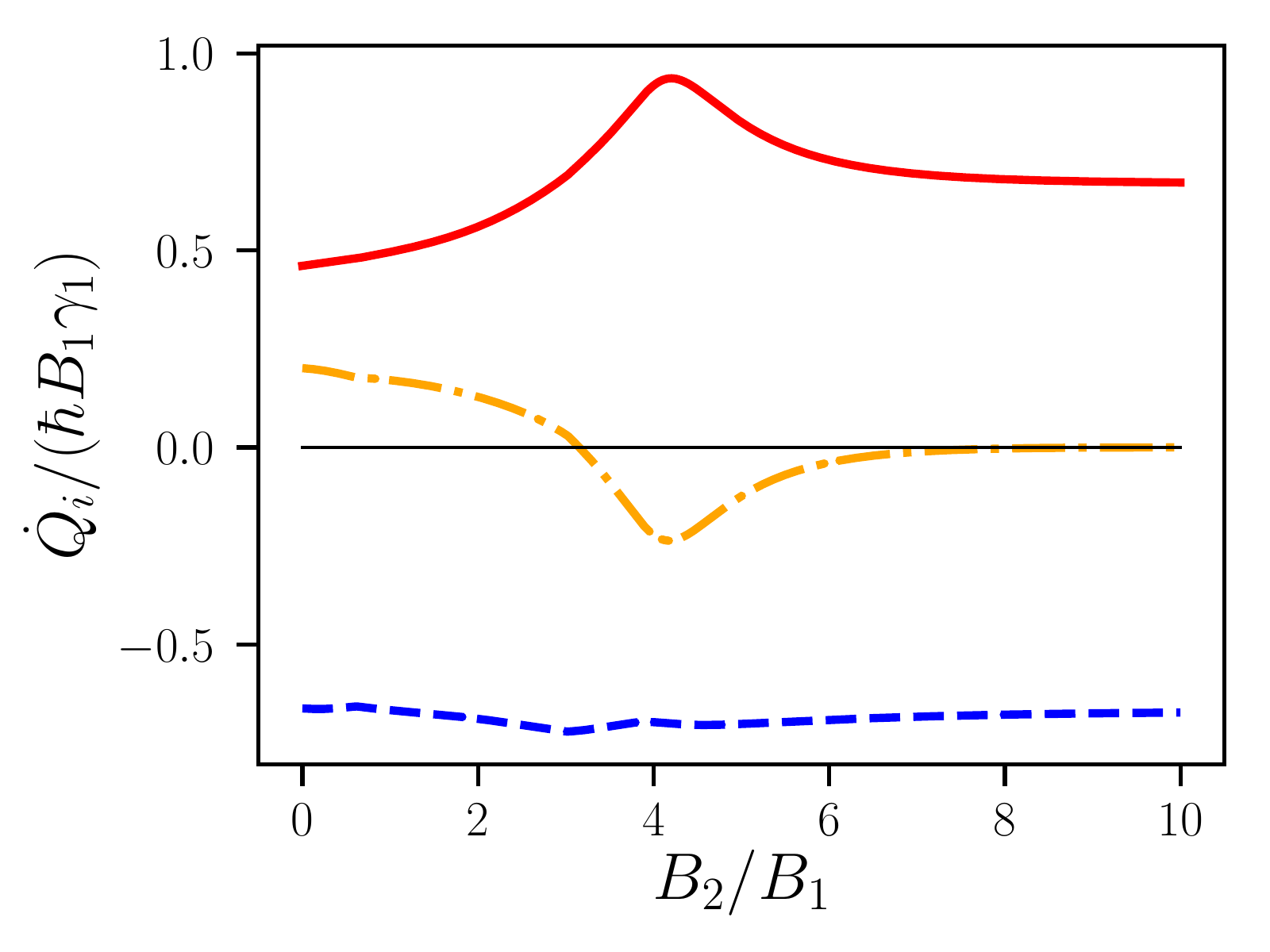}
	\includegraphics[width=0.9\columnwidth]{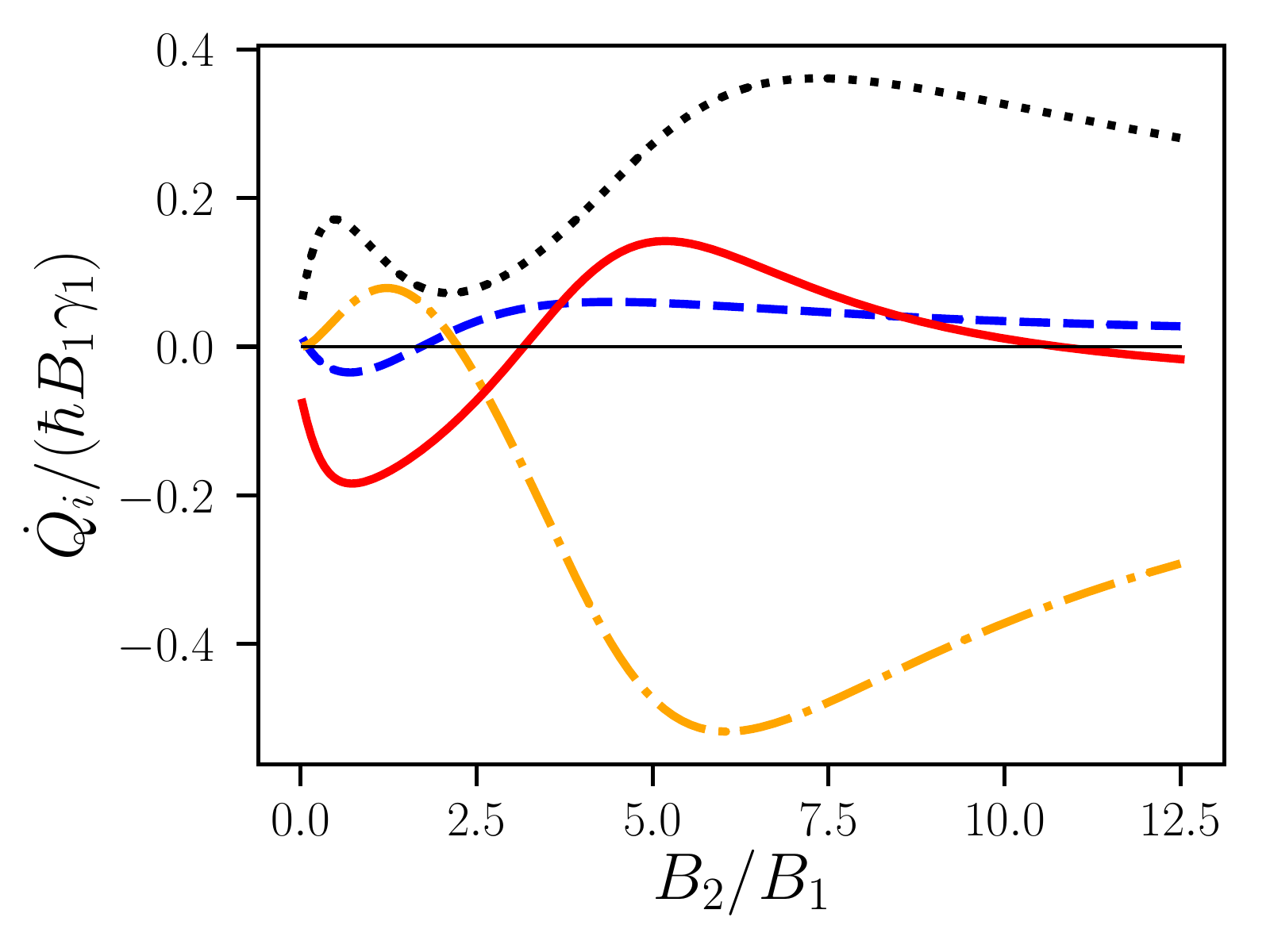}
	\caption{Heat currents $\dot{Q}_1$ (dahsed),  $\dot{Q}_2$ (dot-dahsed), $\dot{Q}_3$ (solid), and the work power $\dot{W}$ (dotted) against the value of $B_2$ in the harmonic-bath (top) and  repeated collision (bottom) models. 
In the harmonic-bath model the control over the currents is even more limited than in the collisional model. For example, in this case the system operates as a heater VIII for small and large values of $B_2$, see Table \ref{fig:operatingmodes}. It operates as a heater X in the transition regime.	
	The set of parameters is $J_{12}=4.07\cdot 10^{-1},J_{13}=3.22\cdot 10^{-1},J_{23}=2.43\cdot 10^{-1},\Delta_{12}=6.31\cdot 10^{-1},\Delta_{13}=7.05\cdot 10^{-1},\Delta_{23}=4.76\cdot 10^{-1},B_1=4\cdot 10^{-1},B_3=1.6, T_1=1,T_2=2,T_3=3$. The bath couplings for the harmonic and repeated interactions models are $\gamma_i=10^{-6}$ and $\gamma_i=10^{-6}$.  }
	\label{fig:heatvalve}
\end{figure}

In this section, we investigate whether our three-qubit setup can be employed as a heat valve in which  qubit 2 is used as a control to regulate the heat current between qubits 1 and 3. In what follows we therefore assume that all the parameters of the device are fixed except for the local magnetic field $B_2$ which will be used as the knob to vary $\dot Q_1$ and $\dot Q_3$.
Our results are summarised in Fig. \ref{fig:heatvalve}. The repeated interactions approach appears to be more flexible for the control of the currents than the harmonic-bath model. This seems to be due to the appearance of the external work cost as an extra energy channel.

%%%%%%%%%%%%%%%%%%%%%%%%%%%%%%%%%
%%%%%%%%%%%%%%%%%%%%%%%%%%%%%%%%%
 We observe three separate heat combinations:
 \begin{enumerate}
\item $\dot{Q}_1<0,~~\dot{Q}_3<0$
\item $\dot{Q}_1>0,~~\dot{Q}_3<0$
\item $\dot{Q}_1>0,~~\dot{Q}_3>0$
\end{enumerate}
with the combination $\dot{Q}_1<0,~~\dot{Q}_3>0$, i.e. heat flowing from 1 to 3, missing. 
This can be explained because we have chosen a set of fields and temperatures such that: $B_1/T_1<B_3/T_3$. According to our diagram in Fig.~\ref{fig:fields}, however, there are no parameters with these assumptions giving heat flow from 1 to 3. Thus although manipulating qubit 2 gives some control over the heat flow between 1 and 3, full control is achieved when the magnetic field of at least two qubits can be changed and the whole diagram in Fig.~\ref{fig:fields} can be explored. We finally remark that, as shown in the bottom  plot of Fig.~\ref{fig:heatvalve}, the heat valve may need external work to function. 

%--------------------------------------------------------------------------------------------------------------------------------
%
%--------------------------------------------------------------------------------------------------------------------------------

%
%--------------------------------------------------------------------------------------------------------------------------------

\section{Conclusions}\label{sec:conclusions}

In this paper we have demonstrated the operating modes of a three-qubit setup as thermodynamic machines. We have shown that it is indeed possible for the three qubits to realise an absorption refrigerator when coupled to three reservoirs at different temperatures. This is true when modelling the system with both the global and local master equation, though the global master equation is more restricted in the accessible operating regimes. Remarkably, this is achieved  assuming simple two-body spin-spin interactions between qubits. Therefore, in contrast to previous designs that require three-body interactions, the scheme we propose could be achieved with current quantum technology setups. 

Besides the absorption refrigerator regime, we have discovered that the three-qubit setup allows a high degree of versatility, operating as several thermodynamic devices.  
Indeed, we have shown how one of the three qubits can be utilised as a heat vale, enabling the control of the heat flow between the other two qubits. 
We stress that this high degree of flexibility requires one to control only local fields and temperatures of the qubits. 

Now that the investigation of three qubit devices has increased our understanding of heat transport in small quantum machines, the next goal is the exploration of mesoscopic setups with many coupled qubits. One important open question is how performance figures such as power, stability and efficiency scale with the number of qubits in the working substance.

\acknowledgements
Some of our  numerical calculations have been performed using QuTiP~\cite{Qutip}.
We gratefully acknowledge financial support by the Spanish MINECO (Grant No. FIS2017-82855-P).
JOG acknowledges an FPU fellowship from the Spanish MECD. GDC acknowledges financial support from the UK EPSRC
(EP/S02994X/1).

%--------------------------------------------------------------------------------------------------------------------------------
%

\appendix
\section{Quantum Correlations}\label{sec:corre}

In this appendix, we examine the relation between the functioning of the three-qubit device and any quantum correlations that exist among the qubits.  As the flow between two qubits is the quantity of interest here we look at correlations between any two qubits as measured by their quantum mutual information:
\begin{equation}
\mathcal{I}_{ij}= S_i+S_j-S_{i,j}, \quad i,j=1,2,3
\end{equation}  
where $S_i$ is the von Neumann entropy $S_i=-{\rm Tr}[\hat{\rho}^i\log \hat{\rho}^i]$ where $\hat{\rho}^i$ is the reduced density matrix of qubit $i$ at steady state. Here we restrict to the repeated interaction model because for the harmonic bath model interqubits currents are zero.

\begin{figure}[t]
	\centering
	\includegraphics[width=0.9\columnwidth]{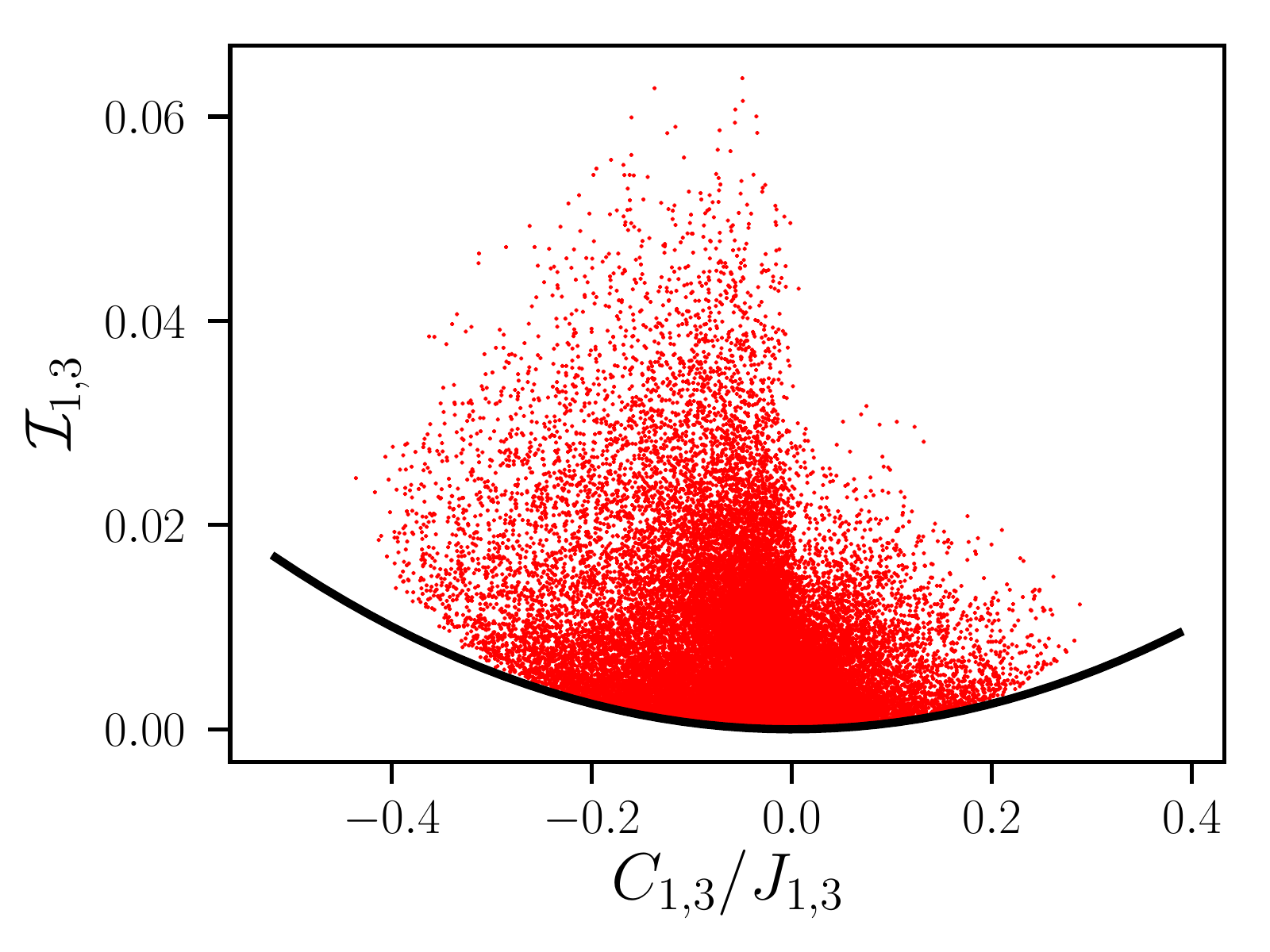}
	\caption{Plot of the mutual information between the qubit 1 and 3 as a function of $C_{1,3}$ for all the points from Fig. \ref{fig:random}.  The black line is the lower bound \eqref{Mutalinformation}.}
	\label{fig:mutualcurrent}
\end{figure}

Plotting the mutual information of a two-qubit pair against their corresponding interqubit current in Fig.~\ref{fig:mutualcurrent}, we see that the values of the mutual information are bounded from below for each value of the interqubit current. This means that to establish a larger current between two qubits these necessarily need to be strongly correlated. 

%%%%%%%%%%%%%%%%%%%%%%%%%%%%%%%%%%
%\begin{figure}[H]
%	\centering
%%	\includegraphics[width=0.85\columnwidth]{Mutual_harmonic_bath.png}
%	\caption{
%Mutual information between the qubit 1 and 3 and $|\rho_{23}^{1,3}|$ for the harmonic-bath model. The refrigerators were obtained with the same parameters than in Fig. \ref{fig:GME_operation_modes}. Notice that for the harmonic-bath model the interqubit currents $C_{ij}\equiv 0$ since  $\dot{W}=0$.
%} \label{fig:GME_mutual}
%\end{figure}
%%%%%%%%%%%%%%%%%%%%%%%%%%%%%%%%%%

%%%%%%%%%%%%%%%%%%%%%%%%%%%%%%%%%
%%%%%%%%%%%%%%%%%%%%%%%%%%%%%%%%%

This fact suggests that there must be a relation between the two quantities as they both depend on the coherences and correlations present in the system. In the following we derive a lower bound for the mutual information in terms of the qubit current. First, in our simulations, we observe that the reduced density matrix for each qubit pair is of the X form:
\begin{equation}\label{2qstate}
\rho_{i,j}=
\begin{pmatrix}
r_{1,1}& 0 & 0 &0 \\
0& r_{2,2} & r_{2,3} & 0 \\
0& r_{2,3}^* & r_{3,3} & 0\\
0& 0 & 0 &r_{4,4}
\end{pmatrix}.
\end{equation}
The interqubit current depends on the off-diagonal element:
\begin{equation}
C_{i,j}=8J_{i,j}{\rm Im}[r_{2,3}].
\end{equation}
The relation between the mutual information and the entries of the reduced density matrix is a little more convoluted. For X states like \eqref{2qstate} the eigenvalues are:
\begin{align}
\lambda_1&= r_{1,1},\\
\lambda_2&=\frac{1}{2}\left[r_{2,2}+r_{3,3}+\sqrt{(r_{2,2}-r_{3,3})^2+4|r_{2,3}|^2}\right],\\
\lambda_3&=\frac{1}{2}\left[r_{2,2}+r_{3,3}-\sqrt{(r_{2,2}-r_{3,3})^2+4|r_{2,3}|^2}\right],\\
\lambda_4&=r_{4,4},
\end{align}
and the mutual information will therefore depend non-linearly on $|r_{2,3}|$. An educated guess for the state with minimum mutual information for a given off-diagonal coherence $|r_{2,3}|$ is a state with equal populations: $r_{m,m}=1/4,\; m=1,2,3,4$. For such a state the mutual information reads simply:
\begin{equation}\label{Mutalinformation}
\mathcal{I}_{ij}=\frac{1}{4}\left(\xi^+\log\xi^++\xi^-\log\xi^-\right),
\end{equation}
where $\xi^\pm=1\pm4|r_{2,3}|$. Therefore the minimum can be achieved by using: $\xi^\pm=1\pm|C_{i,j}|/2J_{i,j}$.

Comparing the lower bound \eqref{Mutalinformation} with the numerical results of the quantum mutual information for qubits 1 and 3 against $C_{1,3}$ in Fig.~\ref{fig:mutualcurrent} we see that the lower bound is indeed very tight.

We have also checked the presence of tripartite entanglement in the system detected by a non positive density matrix after partial transposition. Even for the relatively high values of the temperatures we have considered, we do observe indeed tripartite entanglement which however does not seem to bear a relation with the functioning of the three qubit machine as one of the ten operating modes listed in Table \ref{fig:operatingmodes}.

%--------------------------------------------------------------------------------------------------------------------------------

%--------------------------------------------------------------------------------------------------------------------------------
\bibliography{Refs}
%--------------------------------------------------------------------------------------------------------------------------------
%

\end{document}